# Variational Representational Similarity Analysis (vRSA) for M/EEG

Alex Lepauvre[1, 2], Lucia Melloni[1, 3, 4], Karl Friston[5], Peter Zeidman[5*]

1. Neural Circuits, Consciousness and Cognition Research Group, Max Planck Institute for Empirical Aesthetics, Frankfurt am Main, Germany

2. Chair of Cognitive Computational Neuroscience, Faculty of Psychology, TUD Dresden University of Technology, Dresden 01187, Germany

3. Predictive Brain Department, Research Center One Health Ruhr, University Alliance Ruhr, Faculty of Psychology, Ruhr University Bochum, Bochum, Germany

4. Canadian Institute for Advanced Research (CIFAR), Brain, Mind, and Consciousness Program, Toronto, ON, Canada

5. Functional Imaging Laboratory, Department of Imaging Neuroscience, UCL Queen Square Institute of Neurology, University College London, 12 Queen Square, London. WC1N 3AR.

* Corresponding author

## Abstract

This paper introduces variational representational similarity analysis RSA (vRSA) for electromagnetic recordings of neural responses (e.g., EEG, MEG, ECoG or LFP). Variational RSA is a Bayesian approach for testing whether the similarity of stimuli or experimental conditions is expressed in univariate or multivariate neural recordings. Extending an approach previously introduced in the context of functional MRI, vRSA decomposes the condition-by-condition data covariance matrix into hypothesised effects and observation noise, thereby casting RSA as a covariance component estimation problem. In this context, peristimulus time may be treated as an experimental factor, enabling one to test for the probability that different experimental effects are expressed in data at different times. Variational Bayesian methods are used for model estimation and model comparison, which confer a number of advantages over classical approaches; including statistically efficient hypothesis testing, quantification of uncertainty—using Bayesian credible intervals—and computational efficiency. After introducing the theory, we provide a worked example using openly available EEG data. Software functions implementing vRSA for the SPM software package accompany this paper, together with exemplar analysis scripts.

# Introduction

Here, we consider the analysis of EEG, MEG, OPM or LFP experiments, in which participants are presented with stimuli that have been drawn from different categories or experimental conditions. For example, a "condition" may relate to a specific stimulus that is presented one or more times during the experiment. The aim is to test whether the participants' neural responses distinguished the conditions, and at which times.

This gives rise to three key analysis decisions:

1. Whether to test for experimental effects in one channel or source at a time, or consider patterns that extend over multiple channels or sources together (in a region of interest or across the entire brain). These patterns are sometimes called *representations*.

2. If multiple channels are to be analysed together, whether to analyse them as a matrix of multivariate timeseries, or whether to summarise them using first order or second order statistics. A first order statistic would typically be the mean of the data over channels, and a second order statistic could be their covariance over channels.

3. Whether to specify hypotheses in terms of first order effects (i.e. changes in the level of the measured response due to each experimental condition) or second order effects (i.e. the similarity or difference among experimental conditions).

All permutations of these options can be addressed using the same form of multivariate linear regression model, but with different analysis pipelines and implementation details. Here, our focus is on testing whether experimental effects are expressed in 1) either one or multiple channels, 2) in the *second order statistics* of the data, i.e. the covariance of the conditions, and 3) with hypotheses specified in terms of the *similarity* of the experimental conditions, i.e. the second order statistics of the experimental design. This kind of modelling is described in the neuroimaging literature as *Representational Similarity Analysis* (RSA) (Kriegeskorte, 2008).

In the context of fMRI data, RSA has previously been re-cast in terms of a well-established class of statistical methods known as *covariance components analysis* (Friston, Diedrichsen, et al., 2019). This refers to decomposing a covariance matrix – here the condition-by-condition covariance matrix – into a linear mixture of weighted covariance matrices, called *covariance components*. In the context of RSA, the components or *hypothesis matrices* encode hypothesised contributions to the data. For example, a model might include one covariance component for each experimental condition, additional components for the interactions among experimental conditions, and a component to capture the observation noise. The parameters that weight the contribution of these components – which we will refer to as *hyperparameters* – are estimated from the data. Standard tools are available for covariance component estimation – here we use the Restricted Maximum Likelihood (REML) routine from the SPM software package, which is a *variational Bayes* scheme routinely used in GLM analysis (Friston et al., 2002). Together, this approach to RSA is referred to as variational RSA (vRSA).

Here, we extend vRSA to electromagnetic data (i.e., EEG, MEG, OPM, ECoG or LFP). The approach works for data that are univariate (from a single channel) or multivariate (multiple channels). Its key advantage is that it yields an estimate of the *log evidence*, which is the log of the probability of having seen the data $y$ given a model $m$, $\ln p(y|m)$. Model evidence is also known as the marginal likelihood and can be used to test hypotheses by comparing the evidence for one model against another. In practice, one simply specifies a model $m_1$ that includes the covariance components of interest, and at least one other model $m_2$ where certain components have effectively been "switched off", by fixing their corresponding hyperparameters to zero. The scheme described here furnishes an estimate of the log evidence for each model. Following this, the ratio of evidences (equivalently, the difference in log-evidences) can be reported: $\ln p(y|m_1) - \ln p(y|m_2)$. This is referred to as the *log Bayes factor* (Kass & Raftery, 1995). It is straightforward to convert this result to a posterior probability in favour of either model, $p(m_1|y)$ or $p(m_2|y)$, by applying a softmax function to the log Bayes factors. Comparing models based on their evidence is referred to as *Bayesian model comparison* and can be generalised to compare any number of models. Recent developments of a related technique — *Bayesian model reduction* — enable the log evidence for models with different mixtures of parameters switched on or off to be rapidly assessed, in a matter of milliseconds, without having to separately fit each model to the data (Friston, Parr, et al., 2019).

An important opportunity afforded by EEG/MEG/OPM/ECOG data — which does not typically arise with fMRI data — is the ability to resolve the time within a trial when experimental effects are evident. We will demonstrate the application of vRSA to identify the probability of particular experimental effects being expressed in an EEG dataset at particular times post-stimulus. Our approach is to treat peri-stimulus time as an experimental factor and include covariance components (i.e., hypothesis matrices) encoding the effect of each condition at each peri-stimulus time (more formally, this is the *interaction* of time and experimental condition). Bayesian model comparison is then used to assess how switching covariance components on or off changes the overall log evidence.

We begin by rehearsing the theory underlying vRSA, before establishing its face validity using simulated data, and finally we provide an illustrative application to an openly available EEG dataset. Software implementations that accompany the paper are compatible with the SPM software package.

## Theory

### Generative model

With vRSA, we may be dealing with univariate or multivariate data. The multivariate General Linear Model (GLM) accommodates both:

$$Y = ZU + XB + E \quad (1$$

Where $Y \in \mathbb{R}^{M \times P}$ is the data with $M$ measurements and $P$ measurement channels, the columns of the design matrix $Z \in \mathbb{R}^{M \times V}$ encodes $V$ explanatory variables and the corresponding regression parameters are $U \in \mathbb{R}^{V \times P}$. Nuisance effects – those that are not interesting – are encoded by the design matrix $X \in \mathbb{R}^{M \times W}$ with corresponding parameters $B \in \mathbb{R}^{W \times P}$.

The only distributional assumptions relate to the error matrix $E \in \mathbb{R}^{M \times P}$, which is I.I.D. over measurements within each of the channels, but there may be covariance among channels, defined by the spatial covariance matrix $S \in \mathbb{R}^{P \times P}$.

$$vec(E) \sim N(0, S \otimes I_M) \quad (2$$

The spatial covariance matrix is replicated over measurements by taking the Kronecker product $\otimes$ with the identity matrix of dimension $M$. Details of the definition of the spatial covariance matrix are provided in Appendix A.

### Second order effects

With vRSA we are not interested in the parameters $U$ directly. Rather, we seek to explain the condition-by-condition covariance matrix $G = UU^T$, which encodes the similarity of measurements across experimental conditions. The GLM can be expressed in terms of second order matrices:

$$\Sigma_Y = ZGZ^T + X\Sigma_B X^T + \Sigma_E \quad (3$$

Where $\Sigma_Y = YY^T$ is the covariance of the data, $\Sigma_B = BB^T$ is the covariance of the confounds and $\Sigma_E = EE^T$ is the covariance of the observation noise. Writing the model in this way emphasises that the covariance of the data may be decomposed into a linear sum of terms related to effects of interest, confounds and noise. More specifically, in what follows, we work with the *confound-* *corrected* parameters $\widehat{U}$ covariance and their covariance $\widehat{G}$, which are obtained using the identities provided in Appendix B.

## Covariance component modelling

In vRSA, the confound-corrected condition-by-condition covariance matrix $\widehat{G}$ is decomposed into a linear mixture of covariance components $Q_1, \ldots, Q_n$, also known as hypothesis matrices:

$$\widehat{G} = \widehat{U}\widehat{U}^T = v_1 Q_1 + v_2 Q_2 + \cdots \tag{9}$$

Each of $n$ covariance components can contribute to the overall covariance, weighted by a hyperparameter $v_1 \ldots v_n$, which is estimated from the data. Covariance components could correspond to the individual conditions, or experimental factors, or responses taken from other brain regions or species. In addition to experimentally relevant components, one component is also included to model the covariance of the confound parameters, $B$. This component is specified as $Q_0 = X^+(X^+)^T$ where $X^+$ is the pseudoinverse of $X$.

The covariance components are equivalent to contrasts in a standard GLM analysis. Thus, we can convert from a contrast vector $c_i$ to a covariance component $Q_i$ via:

$$Q_i = c_i c_i^T \tag{10}$$

And from a (rank one) covariance component back to a contrast vector via:

$$c_i = SVD(Q_i) \tag{11}$$

Where $SVD$ is the singular value decomposition, which returns the left singular vector of $Q_i$.

We require the hyperparameters $v_1 \ldots v_n$ to be positive in sign, to ensure that $\widehat{G}$ is a valid (i.e., positive definite) covariance matrix. To ensure positivity, we do not estimate the hyperparameters directly, but rather we estimate latent variables $\lambda = (\lambda_1 \ldots \lambda_n)$, which are the log of the hyperparameters:

$$\lambda_i = \log(v_i) \Leftrightarrow v_i = \exp(\lambda_i) \tag{12}$$

We estimate $\lambda$ using Variational Laplace, which requires defining priors that serve as constraints on the parameters. We use a multivariate normal probability density:

$$P(\lambda) = N(\mu, I_n \cdot \sigma^2) \tag{13}$$

This requires selecting values for the prior mean $\mu$ and prior variance $\sigma^2$. We selected these on the basis of simulations, as described in Appendix C. The optimal values were $\mu = -8, \sigma^2 = 4$, illustrated in Figure 1 (left panel).

Because of the log transform from $\lambda_i$ to $v_i$, this is equivalent to setting a log-normal prior distribution on $v_i$ (Figure 1, right panel). Therefore, whatever value for $\lambda_i$ is selected by the estimation scheme, the corresponding hyperparameter $v_i$ will be positive in sign. For the confound-related component $Q_0$, we use a prior variance of 128 (i.e., a flat prior).

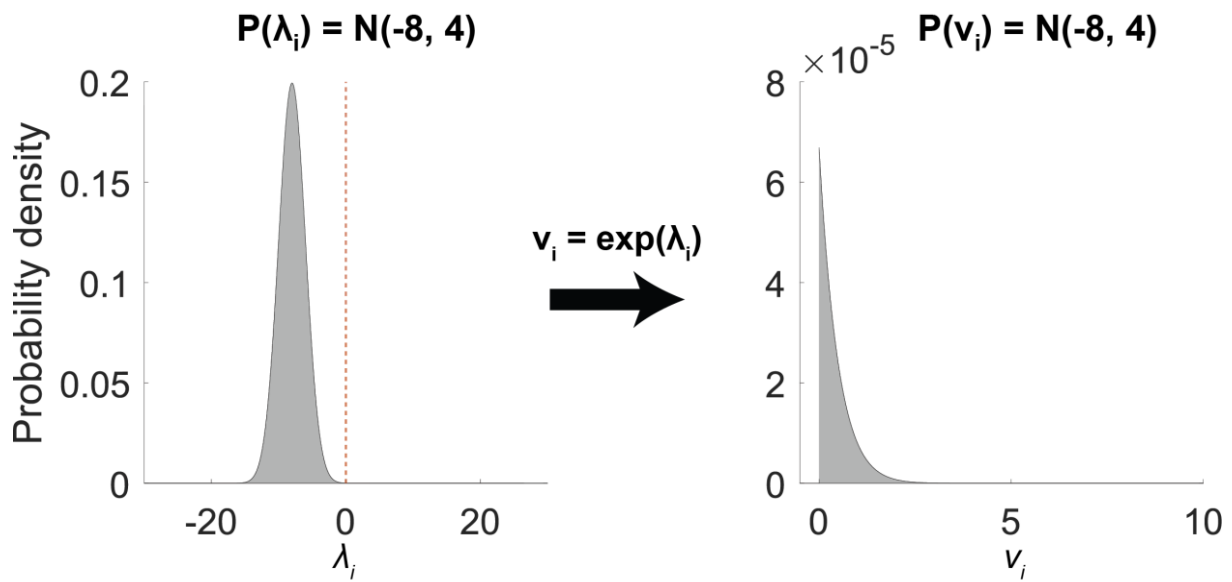

**Figure 1 Priors on hyperparameters. Left:** Normal probability density function that serves as the prior on latent variable $\lambda_i$. This is estimated from the data. Within the model, it is transformed by taking the exponential, ensuring a value that is positive in sign. This is equivalent to setting a log-normal prior on the hyperparameter $v_i$ (right panel).

With vRSA, the parameters $\lambda$ are estimated using the spm_reml_sc function in SPM. This takes as input the covariance matrix to be explained, $\hat{G}$, the confounds $X$, the covariance components $Q_{1...n}$, the spatial degrees of freedom $v_e$ and the definition of the prior density $P(\lambda)$. It returns a posterior probability density over the hyperparameters, $P(\lambda|Y)$, and an approximation of the log evidence which scores the quality of the model. The approximation is called the free energy, $F \approx \ln P(Y|m)$. The free energy — also called the evidence lower bound (ELBO) in machine learning — has the useful property that it can be decomposed into the accuracy minus the complexity of the model. Thus, when comparing models, the model with the largest free energy will offer the best trade-off between being accurate while using as few (effective) parameters as possible.

## Methods

### Data and design

To illustrate the approach we used openly available EEG data from, downloaded from http://purl.stanford.edu/bq914sc3730. Ten subjects viewed 72 images, 72 times per image overall (Figure 2). The content of each image was either animate or inanimate. Within the animate images, there was a 2x2 factorial design: body part (face or body) and species (human or animal). Within the inanimate objects there was a single experimental factor: they were either natural or man-made. Each image was displayed for 500ms, followed by a 750ms grey screen. Subjects were instructed to fixate throughout, while data were collected using a 128-channel EEG system. These data and experimental design allow us to showcase variational RSA by testing for a variety of experimental effects at the group level: for example, is there an experimental effect of faces when accumulating evidence over peristimulus time? Alternatively, is there a particular time point when experimental effects are expressed, when accumulating evidence over stimuli? We will see below that these tests are based upon Bayesian model comparison at the group level, using Bayesian model reduction and Parametric Empirical Bayes.

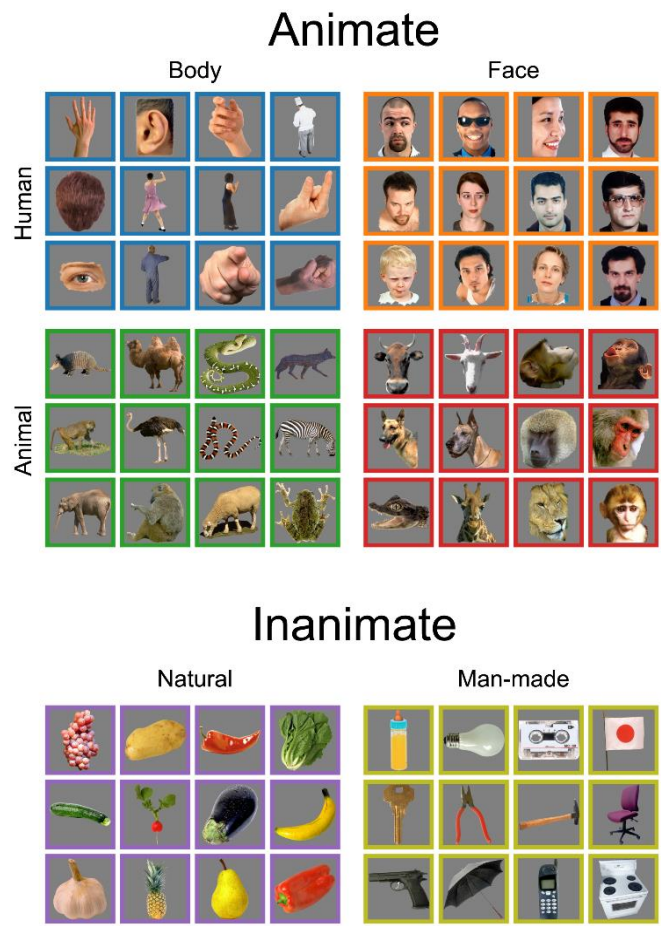

**Figure 2 Stimuli used in the experiment**. Each stimulus had been displayed to the participant 72 times and we averaged the EEG timeseries over repetitions of each stimulus. © 2015 Kaneshiro et al., distributed under the terms of the Creative Commons Attribution License (https://creativecommons.org/licenses/by/4.0/)

Pre-processing

The publicly available EEG data were provided in pre-processed form, having been filtered, downsampled to 62.5Hz and epoched into trials consisting of 32 measurements (496ms post-stimulus) time-locked to the stimulus onset. For full details of the pre-processing, please see (Kaneshiro et al., 2015). We performed two additional pre-processing steps using SPM. We averaged the timeseries over repeated presentations of each stimulus image, then reduced the dimensionality of the data from the original 124 channels to 7 channels (modes) using principal component analysis (PCA) in order to increase signal-to-noise ratio, following standard procedures in SPM for M/EEG.

Design matrix specification

There were $M = 2304 = 31 \times 72$ observations and $P = 7$ channels in total per subject. The rows of the EEG data $Y \in \mathbb{R}^{M \times P}$ were arranged according to EEG measurements $1 \dots 32$ for the first stimulus image (averaged over presentations of this stimulus), then $1 \dots 32$ for the second stimulus image, etc, for all 72 images. The columns of $Y$ pertained to 7 principal modes following PCA. The first mode for every stimulus is illustrated in Figure 3A.

We defined a within-trial design matrix $Z_t \in \mathbb{R}^{32 \times 15}$ to model the timecourse of neural responses within a given trial. This was a Finite Impulse Response (FIR) basis set, with one row for each of the 32 EEG measurements and one column for each of 15 time bins (0.032s per time bin, see Figure 3B). The first two observations per trial (0-0.032s) were left unmodelled to form the implicit baseline for the model, and the 15 time bins spanned the remaining duration. Each time bin spanned two EEG measurements.

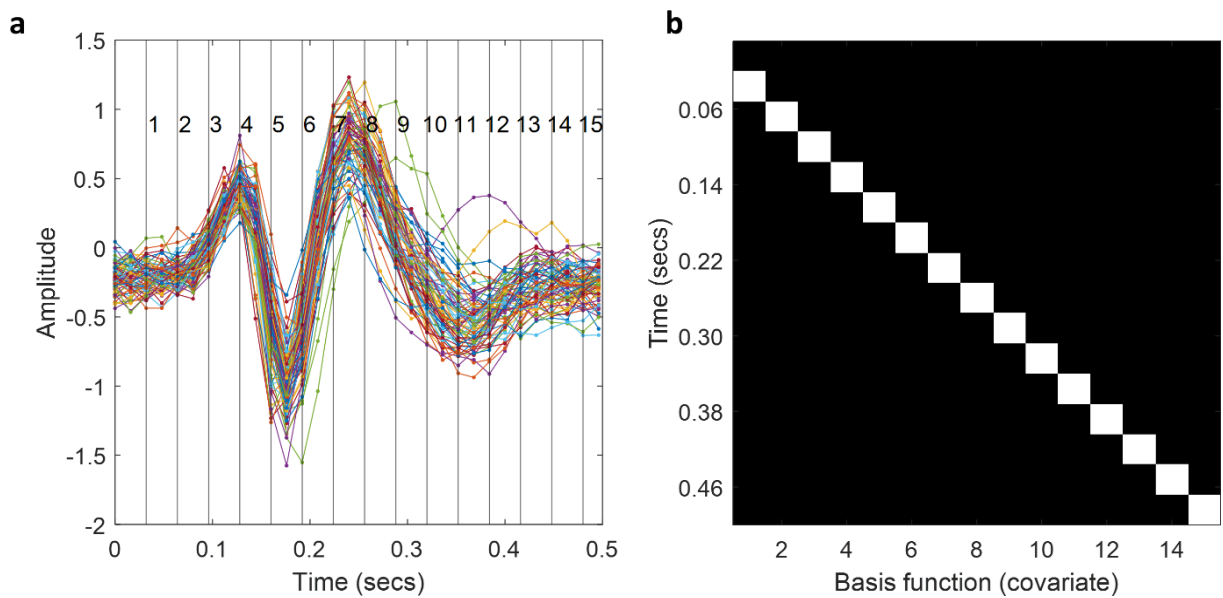

**Figure 3 Example EEG data and within-trial design matrix. a**. The first principal component (i.e., mode or eigenvector) of an example subject's EEG time course. Each time series corresponds to an experimental

condition (i.e., 72 images, averaged over trials). The numbers above the EEG data indicate the 15 Finite Impulse Response (FIR) time bins (covering the entire trial duration beyond the left-out baseline period) into which the data were divided. **b.** Within-trial design matrix $X_t$. The vertical axis corresponds to the peristimulus time within each trial and the horizontal axis are the 15 FIR time bins. White=1, Black=0.

The software routine accompanying this paper automatically handles the process of duplicating the within-trial design matrix (Figure 3b) over the $K = 72$ stimulus conditions, to form the overall design matrix $Z = I_K \otimes Z_t$ (Figure 4).

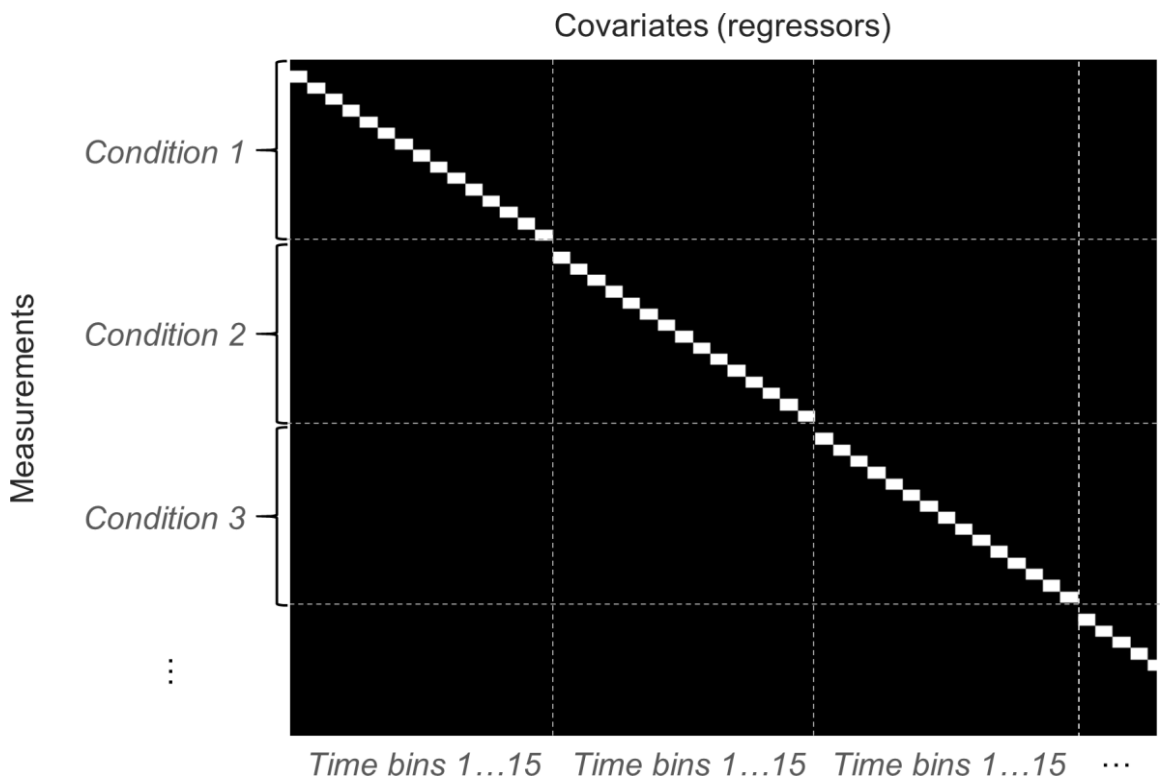

**Figure 4 Overall design matrix $Z$.** The design matrix has dimension $Z \in \mathbb{R}^{M \times V}$, where the rows correspond to the $M = 2304$ EEG measurements and the columns are $V = 1080$ covariates (a covariate per time bin per condition). Here an experimental "condition" is the average EEG data over repeated presentations of a particular stimulus. The rows are ordered according to all the measurements from condition 1, then all the measurements from condition 2, etc. In this example, each condition corresponds to a particular image in Figure 1. For illustration purposes, the matrix is truncated with only the top-left corner shown.

For this example, the confounds design matrix consisted only of the overall mean of the signal, thus a column of ones: $X = \mathbf{1}_M$.

### Contrasts and covariance components

Having specified our explanatory variables — i.e., stimulus and time-specific responses — we can now specify hypotheses in terms of contrasts over the columns of the ensuing design matrix. We first specified five contrast vectors each of length 72 – corresponding to the 72 conditions (stimulus images) – expressing the experimental design: (1) animate-inanimate, (2) species: human-animal,

(3) body part: body-face, (4) natural-manmade, (5) interaction of species and body part. These were specified using 1s and -1s, which were then mean-centred (Figure 5). The software routine then replicates these contrast vectors over the 15 time bins (basis functions) per trial, resulting in five contrasts per time bin. Each contrast was converted to a covariance component (i.e., hypothesis matrix) according to Eq. 10 and a final covariance component was included to model noise. The resulting 76 = 5 × 15 + 1 components entered the model estimation scheme.

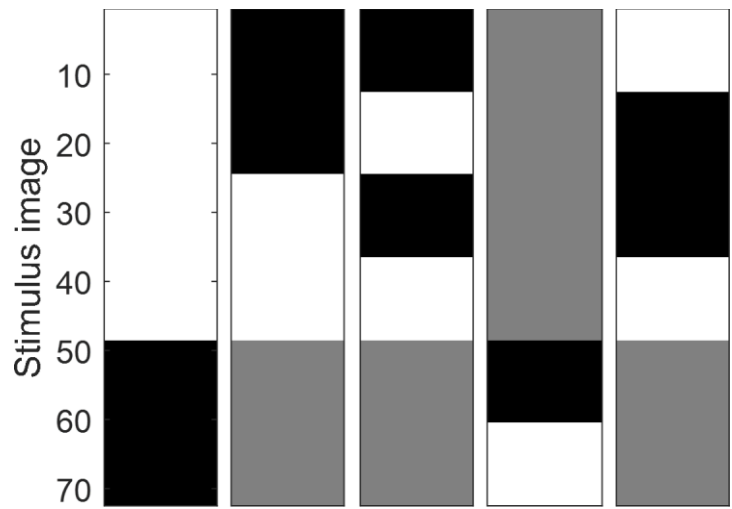

**Figure 5 Contrast specification.** Contrast vectors encoding the experimental conditions. From left to right, these were: animate-inanimate, human-animal, body-face, natural-manmade, interaction of species and body part. White = 1, black = -1, grey=0.

### vRSA model fitting

The analysis begins by estimating the regression parameters, using the standard formula for a GLM:

$$U = Z^+ Y$$

Where $Z^+$ is the pseudoinverse of $Z$. It then converted $U$ to the condition-by-condition covariance matrix $G$, where for this experiment, a "condition" was a particular stimulus at a particular peri-stimulus time:

$$G = UU^T$$

The analysis then uses Variational Laplace (i.e., the REML algorithm in SPM) to decompose the confound-corrected covariance matrix $\hat{G}$ into the 76 covariance components specified above:

$$\hat{G} = \lambda_1 Q_1 + \lambda_2 Q_2 + \cdots + \lambda_{76} Q_{76}$$

### Evidence for experimental effects

To ensure that the models capture experiment-related effects, we tested whether they explained the data better than reduced models that included only the noise component. To do, this we specified an alternate model per subject, where all the covariance components (except for the noise component) were effectively "switched off", by fixing their corresponding hyperparameters to zero. In more detail, the prior probability density for the $i$-th parameter to be switched off was set to $P(\lambda_i) = N(\mu_i, 0)$, thereby fixing that parameter to its prior mean. The change in free energy due to modifying the prior was calculated analytically using Bayesian model reduction (implemented in the SPM function spm_log_evidence_reduce). The difference in free energy (the log Bayes factor) then quantifies the evidence in favour of the model that included experimental effects.

### Group analysis using PEB

To identify effects over subjects, we used standard tools in SPM for Bayesian hierarchical linear regression, called Parametric Empirical Bayes (PEB) (Friston et al., 2016, p. 20; Kass & Steffey, 1989; Zeidman et al., 2019). Using PEB, we specified a hierarchical model comprising the vRSA models at the individual-subject level and a general linear model (GLM) at the group level. Assembling all of the subjects' hyperparameters into a single vector $\boldsymbol{\lambda}_g$, the PEB model was defined as:

$$\boldsymbol{\lambda}_g = \boldsymbol{W\theta} + \boldsymbol{E^{(2)}}$$
$$\widehat{\boldsymbol{G}}^{(s)} = \exp\left(\lambda_1^{(s)}\right)\boldsymbol{Q_1} + \exp\left(\lambda_2^{(s)}\right)\boldsymbol{Q_2} + \cdots$$

The second line of the equation says that $\widehat{\boldsymbol{G}}^{(s)}$ — the condition-by-condition covariance matrix from subject $s$ — was modelled according to the covariance component model introduced earlier. The first line says that a vector containing the hyperparameters from all subjects $\boldsymbol{\lambda}_g = \left(\lambda^{(1)}, \lambda^{(2)}, \ldots\right)$ were modelled by a GLM with between-subjects design matrix $\boldsymbol{W}$. The corresponding regression parameters $\boldsymbol{\theta}$ encoded the group average hyperparameters, as well as the effects of any between-subjects covariates (none was included here). Error matrix $\boldsymbol{E^{(2)}}$ encoded unexplained between-subject variability (i.e., random effects).

Here, we took the 75 hyperparameters quantifying the effect of each contrast at each time to the group level. The group-level design matrix $\boldsymbol{W}$ therefore had 75 columns – encoding the group average of each hyperparameter. Fitting this PEB model to the data returned a posterior probability density over the group-level parameters, $\boldsymbol{\theta}$, as well as the free energy approximation of the log evidence, which scored the quality of the complete hierarchical model.

We then performed a series of Bayesian model comparisons; comparing the log evidence for the full PEB model against reduced models in which particular mixtures of the parameters $\theta$ were switched off. First, we wanted to identify experimental effects at the group level by suppressing any parameters that did not contribute to the model evidence. To this end, we used standard tools in the SPM software to perform an automatic Bayesian model comparison, which evaluates large numbers of reduced models with mixtures of parameters switched off (fixed at their prior expectation). This was performed rapidly and automatically using Bayesian model reduction. The parameters of the best models from this search over models were then averaged (*Bayesian model averaging*). It is these averaged parameters that we report.

Next, we quantified the evidence for each experimental contrast (collapsing over post-stimulus time). Starting again with the full PEB model, we turned off all the parameters relating to a particular experimental condition, and we repeated this for each condition. For example, we turned off the 15 parameters encoding the effects of 'Face-Body' (one parameter per peristimulus time bin). We recorded the change in free energy and then started again for each of the remaining four experimental conditions.

Similarly, we quantified the evidence that each of the 15 time bins showed experimental effects (collapsed over contrasts). Starting again with the full PEB model, we turned off all components relating to each time bin in turn. For example, we turned off the five components encoding the effects of all contrasts on time bin 1. We recorded the change in free energy and then repeated this for the remaining 14 time bins.

We report the results of Bayesian model comparisons in two ways. The log Bayes factor is simply the difference between a model's log evidence and the log evidence of some reference model (approximated by the difference in free energies). A log Bayes factor of three equates to $\exp(3) \approx 20$ times the evidence in favour of one model or another, and is regarded as "strong evidence" (Kass & Raftery, 1995). We also convert the log Bayes factors to probabilities for each model given the data, which under equal priors for all models is a softmax function of the log evidences. In what follows, we apply the above procedures to simulated data — to establish model identifiability and face validity — and then turn to the empirical data.

## Results

### Simulation results

Before applying the procedure described above to empirical data, we first performed simulations to confirm that the model could recover the presence of known effects (i.e., face validation). We simulated multivariate data for 10 virtual subjects, using the same design matrix and contrasts as for the empirical analysis that follows. For these simulations, we expressed two experimental effects in the data: the two-way interaction of Species and Body-part in the fourth time bin, and the main effect of Animate-Inanimate in the sixth time bin. There was therefore an interaction between peristimulus time and condition. The effect sizes and amplitude of observation noise were based upon the posterior estimates following analyses of the empirical data described later. We fitted a vRSA model with 76 covariance components corresponding to the effects of each contrast on each time bin, and finally one component modelling observation noise. We expected to find positive evidence in favour of the two covariance components encoding the effects that were present in the simulated data, and negative evidence (i.e., evidence in favour of the null) for all other effects.

The free energy for each virtual subject's full model, relative to a reduced model with only a noise component, ranged from 749 to 757 across all subjects. Converting to posterior probabilities, this was equivalent to a probability of unity in favour of the full model for every subject. Therefore, we could be confident that the models were detecting experimental effects in the data. Next, we examined the estimated model parameters and used them to test hypotheses.

The group-average of the hyperparameters $\lambda$ governing the contribution of the covariance components was estimated using the Parametric Empirical Bayes (PEB) framework. Figure 6a shows these group-average parameters, following an automatic search to prune any parameters not contributing to the model evidence. Most parameters were switched off, i.e. fixed at their prior expectation, with the exception of the two parameters where real effects were present in the simulated data. Figure 6b shows the exponential of these parameters, i.e., the estimated hyperparameters $\nu$.

Recall that we set out to address two questions: which experimental conditions were expressed in the data, and when were they expressed. To validate that vRSA can be used to address these questions, i.e., that it correctly assigns a probability to each possible outcome, we performed Bayesian model comparisons using the PEB model.

The evidence in favour of each covariance component (i.e. each contrast at each time point) being included in the model compared to being fixed at zero is shown in Figure 6c. Each value is the log Bayes factor in favour of a model that includes all hyperparameters against a model with the indicated parameter being switched off. Negative values indicate evidence in favour of the null. As expected, contrasts 1 and 5 (species vs. body; animate vs. inanimate) were detected in the simulated data at time points 4 and 6 (0.12–0.14s and 0.19–0.21s), with log evidence values of 12.54 and 20.45, respectively, confirming the sensitivity of vRSA. For contrasts 2–4, log evidence ranged from – 0.11 to –0.97 across all time points, confirming the absence of these effects and demonstrating specificity.

Figure 6d plots the log evidence (blue line) and posterior probability (grey bars) at each time bin across all contrasts. This correctly identified effects as occurring with ceiling probability in time bins four and six. At all other time bins, log evidence ranged from –2.22 to –4.11, indicating evidence for the absence of effects. Figure 6e shows the log evidence (left) and posterior probability for each contrast across all time points. Strong evidence supported contrasts 1 and 5 (log evidence = 11.02 and 5.01), while contrasts 2–4 showed log evidence between –9.44 and –9.98, reflecting strong evidence against their presence. In summary, when vRSA was applied to data from a simulated group of subjects, it performed as expected and correctly identified the presence and absence of experimental effects. We next applied the same analysis procedure to the empirical data.

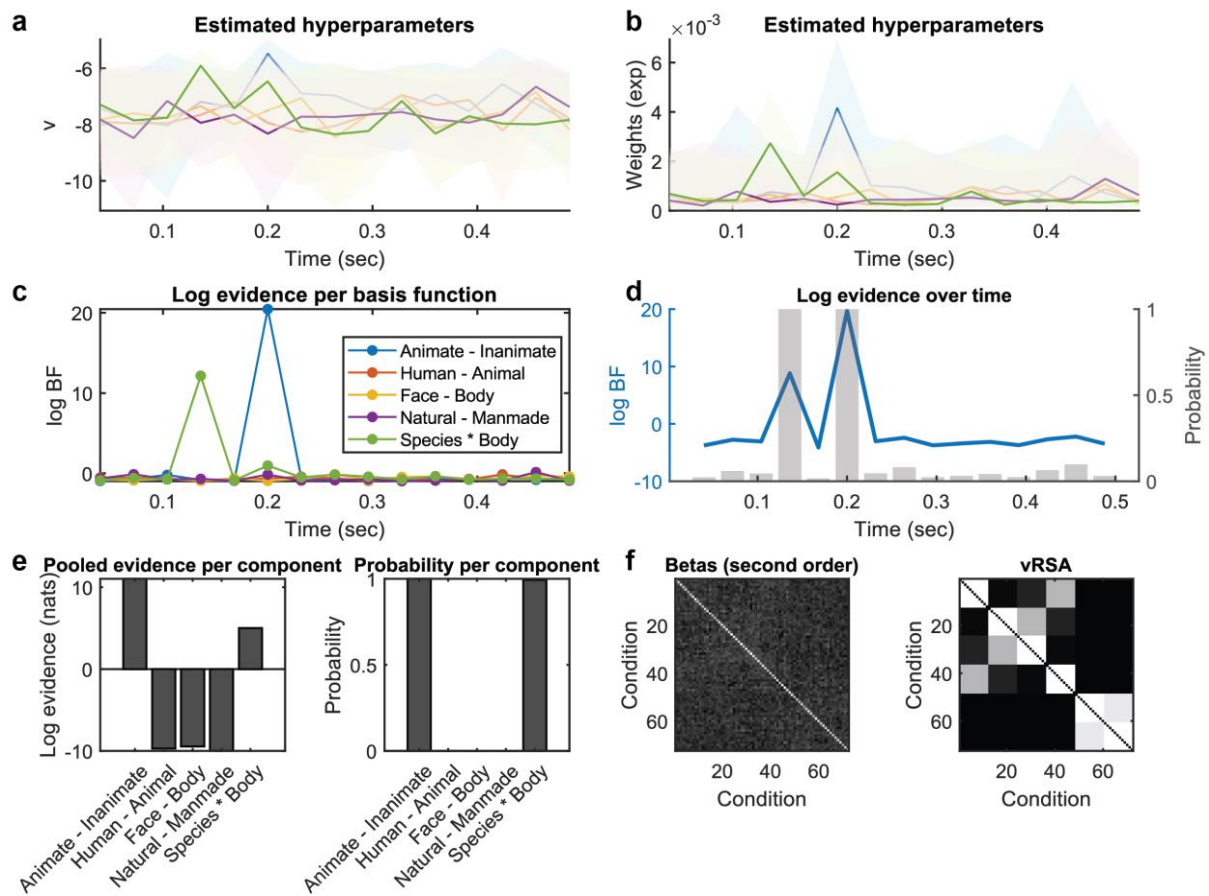

**Figure 6 Simulation results**. **a.** Estimated parameters following automatic search and averaging of PEB models. Each time series represents the group average value of a hyperparameter for a given contrast over time (see panel c for legend). Note that these are log-scaling parameters, which are 'un-logged' by taking their exponential before multiplying them by the corresponding covariance component. Shaded areas are 90% credible intervals. **b.** The same parameters as in part **a** after taking the exponential of their expected values. **c.** Log Bayes factor for each component at each time point, estimated by comparing the model where the component at a given time point is switched on vs. off.  **d**. (**Left y axis and blue time series**) Log Bayes factor over time across conditions, estimated by summing the evidence for each experimental condition at a given time point. (**Right y axis and grey bar plot**) Posterior probability of the presence of second order effects at each time point estimated using

softmax function on the free energy. **e. (Left)** Log Bayes for each condition across time, estimated by summing the evidence for each experimental condition across time. **(Right)** Corresponding posterior probabilities. **f. (Left)** Betas covariance matrix summed across time points with posterior probability >0.9. **(Right)** Weighted sum of the model covariance components.

### Within-subject log evidence

For each subject's empirical data, we fitted a vRSA model with the same 96 covariance components as for the simulation above. The free energy for each subject's full model, relative to a reduced model with only a noise component, ranged from 1.46 to 23.41 across all subjects. Converting to posterior probabilities, this was equivalent to a probability of unity in favour of the full model for every subject. Thus, we could be confident that the models were detecting experimental effects in the data. Next, we examined the estimated model parameters and used them to test hypotheses.

### Group analysis: parameters

We used PEB to perform group level analysis. First, we performed an automatic search that removed any parameters from the PEB model that did not contribute to the overall (group level) free energy. The resulting parameters are shown in Figure 7a. Most parameters deviated from their prior expectation of around -4. Figure 7b shows the parameters more clearly after taking the exponential of their expected values. Three parameters were particularly strong: the parameter encoding the effect of animate vs inanimate stimuli in FIR bin 5 (0.16-0.18s), the effect of faces vs body part in FIR bin 5 (0.16-0.18s) and the interaction of species (human vs animal) and body part (face vs body) in FIR bin 6 (0.19-0.21s). This interaction means that the difference in response to faces and bodies depended on whether humans or animals were depicted.

### Bayesian model comparison

With this analysis, we set out to address two questions: which experimental conditions were expressed in the data, and when were they expressed? To formally address these questions — i.e., to assign a probability to each possible outcome — we performed Bayesian model comparisons using the PEB model.

The evidence in favour of each experimental effect, relative to being fixed near zero, is shown in Figure 7c. The y-axis indicates the difference in free energy for the presence of all relevant parameters versus their absence and the colour indicates the experimental condition. In line with the estimated parameters, we observed strong evidence for the effect of animate vs inanimate stimuli in FIR bin 5 (0.16-0.18s), the effect of faces vs body part in FIR bin 5 (0.16-0.18s) and the interaction of species (human vs animal) and body part (face vs body) in FIR bin 6 (0.19-0.21s). Figure 7d further shows strong evidence from time bin 5 through 8 (from 0.16 to 0.27s) across conditions. Converting to posterior probability (Figure 7d, grey bar), the posterior probability for time bin 5 through 8 was unity. Finally, figure 7e (left) shows strong evidence for all three contrasts (animate vs inanimate, face vs body part and interaction between species and body part) with unity of posterior probability.

In summary, we applied the vRSA to identify which experimental effects were expressed in the data, and when these effects were evident. We can conclude that the strongest effect was the species by body part interaction (probability 1), and the strongest expression of experimental effects was in time bin 6 (probability 1).

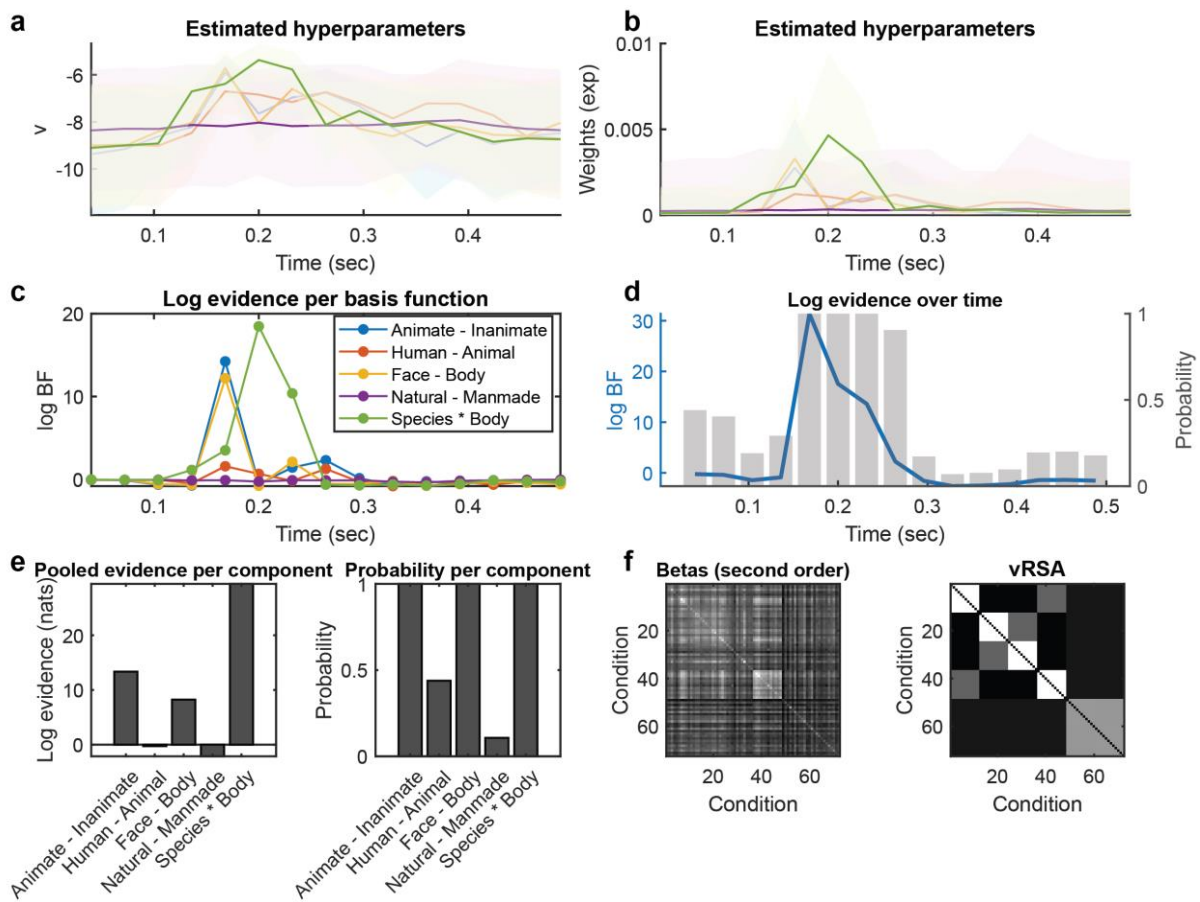

**Figure 7 Empirical results**. **a.** Estimated parameters following automatic search and averaging of PEB models. Each time series represents the group average value of a hyperparameter for a given contrasts over time, pertaining to one experimental condition at one time (see figure c for legends). Note that these are log-scaling parameters, where their exponential is taken before multiplying them by the corresponding covariance component. Shaded areas are 90% credible intervals. **b.** The same parameters as in part **a** after taking the exponential of their expected values. **c.** Log bayes factor for each component at each time point, estimated by comparing the model where the condition at a given time point is switched on vs. off. **d**. (**left y axis and blue time series**) Log Bayes factor over time across conditions estimated by summing the evidence for each experimental condition at a given time point. (**right y axis and grey bar plot**) Posterior probability of the presence of second order effects at each time point estimated using softmax function on the free energy. **e. (Left)** Log Bayes for each condition across time, estimated by summing the evidence for each experimental condition across time. **(Right)** Corresponding posterior probabilities. **f. (Left)** Betas covariance matrix summed across time points with posterior probability >0.9. **(Right)** Weighted sum of the model covariance components.

## Discussion

This paper introduced variational representational similarity analysis (vRSA) for electromagnetic recordings of neural responses. The condition-by-condition covariance matrix of estimated responses is decomposed into a weighted mixture of covariance components, which encode the contributions of hypothesised effects and observation noise. Hypothesis testing at the single subject or group level is made straightforward through the use of Bayesian model comparison. This approach can be applied to univariate or multivariate data without modifications to the code. Software and example scripts are provided with this paper to reproduce the analyses presented above.

vRSA has several strengths over alternatives, including certain standard RSA analyses, linear classifiers and multivariate pattern analysis (MVPA). First, a major advantage is that it enables the full experimental design – i.e., both main effects and interactions – to be properly represented in the model and distinguished from observation noise. This is not the case for RSA models or classifiers that only consider one experimental condition at a time. Second, vRSA is a variational Bayes approach that provides an estimate of the log model evidence. This is the crucial quantity needed for comparing any number of models, which differ in the choice of covariance components. The model with the highest log evidence or free energy offers the best trade-off between accuracy and complexity among those compared. Third, by the Neyman-Pearson lemma, hypothesis testing based on the approach used here is statistically optimal, and should have equal or greater statistical power compared to hypothesis tests based on other quantities, such as classification accuracy. Fourth, the use of variational Bayes and Bayesian model reduction — which apply numerical approximations and analytic solutions in place of sampling — make model estimation and comparison extremely fast on a standard desktop computer and exactly reproducible. And fifth, the PEB framework makes it straightforward to generalise RSA analysis from individual subjects to groups, in order to investigate both the commonalities and differences among participants.

As noted when vRSA was first introduced (Friston, Diedrichsen, et al., 2019), there is a close correspondence between vRSA and another approach called Pattern Component Modelling (PCM) (Diedrichsen et al., 2018). Both methods share the same generative model – a multivariate linear model with a covariance component formulation. A key distinction is the use of variational Bayes in vRSA, which avoids the use of cross-validation. Nevertheless, in principle, PCM could also be applied to handle peri-stimulus time for M/EEG analysis as described here, and we would expect both approaches to give similar results.

The key limitation of vRSA — and of the application of GLM analysis to neuroimaging data more generally — is that it provides a descriptive rather than a mechanistic model of the data. It can be used to test which experimental effects are present or absent, and as shown here, the time at which those effects occur. However, vRSA offers no insight into the physiological genesis of the data. Neurobiologically plausible (e.g., neural mass) models provide a viable alternative, where parameters such as synaptic time constants have a clear biological interpretation. The Dynamic Causal Modelling (DCM) framework in SPM provides the necessary tools for fitting neural mass models to EEG/MEG/LFP data. This could be used, for example, to investigate whether differences in the response to certain stimuli apparent in the data (Figure 3A) could be explained by differing effective connectivity among relevant brain regions, such as face-selective fusiform gyrus. Thus, vRSA may serve a useful function in terms of identifying where and when experimental effects are expressed for subsequent dynamic causal modelling — with greater statistical efficiency than other multivariate analysis methods as described above.

A criticism that may be levelled against vRSA, and RSA more generally, is that modelling the covariance of the data introduces an unnecessarily complicated analysis procedure. Any hypothesis matrix, encoding the similarity or dissimilarity of experimental conditions, could be transformed to a contrast vector or matrix (Eq 11) and used to test a linear contrast of the parameters in a multivariate GLM directly (Eq 1). The standard approach for linear regression with multivariate data using classical frequentist statistics is called MANOVA, which is typically combined with canonical correlation analysis (CCA) or canonical variates analysis (CVA) to identify pairs of weights over the data channels and design that best explain the data. For Bayesian analysis, a univariate linear regression model is provided in various analysis packages. In SPM, this is available in the function spm_peb. To analyse multivariate data using a univariate GLM, the data can be vectorized (the channels concatenated), with appropriate covariance components defined to enable differing levels of observation noise per channel. Nevertheless, RSA has gained popularity in the fMRI community (Haynes, 2015; Kriegeskorte & Kievit, 2013; Poldrack et al., 2009). This is partly because of the intuitive appeal of talking about the similarity or dissimilarity of measurements and hypotheses, and partly because it is straightforward to include similarity or dissimilarity matrices from different sources; for example from different species, brain regions or imaging modalities (Cichy et al., 2014; Kriegeskorte et al., 2008). The methods presented here enable the same applications, with the added benefits that come from Bayesian inference.

The empirical analysis we presented provides a straightforward illustration of the approach using the kind of data typically analysed with RSA. However, there are opportunities for further optimisation, particularly with regards to the specification of the design matrix $Z$. Here, we used an FIR model, which meant having one column in the design matrix per time bin (15 in total here). The efficiency of the model could be increased by having a design matrix with fewer columns, thereby increasing the available degrees of freedom of the model. If the aim of the analysis is to test theory-specific predictions regarding the temporal dynamics underlying the neural representations, the design matrix can be specified to reflect these hypothesized time courses. Alternatively, one approach would be to replace the FIR model with a small number of basis functions which, when summed, produce the shape of an ERP. This is how fMRI data are commonly analysed, with one basis function encoding a canonical haemodynamic response function, and one or two further basis functions (referred to a temporal and spatial derivatives) that can shift the peak and width of the modelled haemodynamic response.

Readers wishing to apply the methods described here may wish to begin with the example analysis presented above. A script to download the data and perform the analyses accompanies this paper. More broadly, vRSA illustrates how Bayesian inference can unify representational and model-based approaches in cognitive neuroscience. By framing representational similarity as a problem of covariance component estimation, vRSA provides a statistically optimal quantification of evidence for representational models. Beyond EEG and MEG, this framework could be extended to cross-species, cross-modality, or longitudinal studies, where questions about when and how representations emerge are central. In this sense, vRSA is not only a tool for efficient analysis, but a step toward a more integrated and transparent understanding of neural representations across methods and levels of explanation.

## Appendix A: Spatial degrees of freedom

This appendix explains the definition of spatial covariance matrix $\boldsymbol{S}$. To briefly rehearse an aspect of multivariate statistics, for any matrix $\boldsymbol{A} \in \mathbb{R}^{m \times n}$, where each column $i$ is independently sampled from a multivariate normal density with mean zero and covariance matrix $\boldsymbol{\Sigma}$, i.e., $\boldsymbol{A_i} \sim N(\boldsymbol{0}, \boldsymbol{\Sigma})$, the summed covariance of all columns has a Wishart distribution. This is the multivariate generalization of the gamma distribution, and is written as:

$$\sum_{i=1}^{n} A_i A_i^T \sim W_m(\boldsymbol{\Sigma}, n) \tag{A.1}$$

Where the first parameter of the distribution is called the scale matrix and the second is called the degrees of freedom. In the context of a multivariate linear regression model, the error covariance matrix $\boldsymbol{EE}^T$ follows a Wishart distribution:

$$\boldsymbol{EE}^T \sim W_M(\boldsymbol{I_M}, \nu_e) \tag{A.2}$$

Here, we set the temporal error covariance to the identity matrix $I_M$, thereby making an I.I.D. approximation over time. If every measurement channel were independent, then the spatial degrees of freedom would equal the number of measurement channels, $v_e = P$. However, to accommodate spatial covariance, we set the degrees of freedom based on the spatial covariance matrix $S$, according to the following estimator (Seber & Lee, 2003; Worsley & Friston, 1995):

$$v_e = \frac{tr(S)tr(S)}{tr(SS)} \tag{A.3}$$

This is the variance due to the channels independently, over the total variance including their covariance. As the spatial covariance increases, the spatial degrees of freedom decrease. This enters into the covariance component estimation scheme used in SPM (spm_reml_sc.m).

## Appendix B: Confound-corrected covariance matrix

To obtain matrix $G$ after correcting for known confounding effects encoded in $XB$, first consider the univariate linear model:

$$y = zu + xb + e \tag{4}$$

To isolate the parameters relating to the experimental design $u$, we have:

$$\frac{y}{z} = u + \frac{xb}{z} + \frac{e}{z} \tag{5}$$

In matrix form, rather than dividing each term by $Z$, we pre-multiply by its generalized inverse, $Z^+$:

$$\begin{aligned} Z^+Y &= Z^+ZU + Z^+XB + Z^+E \\ &= U + Z^+XB + Z^+E \end{aligned} \tag{6}$$

The confounds are expressed in the term $Z^+X$. To remove these effects, we define the *residual maker matrix* or *residual forming matrix* $R = R^T$:

$$R = I - (Z^+X)(Z^+X)^+ \tag{7}$$

We pre-multiply this by all terms, and refer to the resulting expression as $\widehat{U}$:

$$\widehat{U} := RZ^+Y = RU + RZ^+E \tag{8}$$

The confound-corrected parameters $\widehat{U}$ can be decomposed into the effects of the task plus the observation noise. This serves as the basis for calculating the condition-by-condition covariance matrix, corrected for confounds, $\widehat{G} = \widehat{U}\widehat{U}^T$. This covariance matrix can be decomposed into a sum of terms that depend on the (confound-corrected) covariance of experimental effects plus the covariance of the noise.

## Appendix C: Selection of Priors

To recap, the parameters that weight the covariance components $v = v_1, \ldots, v_n$ are not estimated directly. To ensure that they are positive in sign, latent variables $\lambda = \lambda_1, \ldots, \lambda_n$ are estimated instead, which are related to the model parameters according to $v_i = \exp(\lambda_i)$. Because the exponential operator always returns a positive number, we can guarantee positive values of $v_i$, regardless of the value chosen by the model fitting algorithm for $\lambda_i$.

We chose a normal distribution for the prior on the latent variable, $p(\lambda_i) = N(\mu, \sigma^2)$, which corresponds to placing a log-normal prior on the actual parameter $v_i$, thus $p(v_i) = \text{Lognormal}(\mu, \sigma^2)$. We wanted to have a prior expectation on $v_i$ close to zero, expressing the assumption that there are no experimental effects unless the data proved otherwise. To achieve this, we needed to set a prior expectation with negative sign on $\lambda_i$. For example, a prior expectation of $\lambda_i = -8$ corresponds to $v_i = \exp(-8) = 3.54e^{-4}$.

The subtlety is how to choose the best value for the prior expectation, $\mu$, and the prior variance $\sigma^2$. If $\mu$ is too negative, the prior will be concentrated around $v_i = 0$ and preclude values that are large enough to explain the data (unless $\sigma^2$ is very large). Also, because large deviations from the prior on $v_i$ correspond to very small changes in the $\lambda_i$, this results in small effects being assigned a very high posterior probability, which we found for the essentially flat prior of $\mu = -16, \sigma^2 = 128$. Alternatively, if $\mu$ is too positive, then 'switching off' covariance components requires a large departure from the prior, leading to a penalty for simpler models. We therefore need to choose an appropriate prior mean and variance, to balance sensitivity and efficient regularization of parameters.

The optimal combination of $\mu$ and $\sigma^2$ depends on the signal to noise ratio (SNR) of the data. However, time-resolved modalities (such as EEG, MEG, and iEEG) typically have different SNR, making a single default choice difficult. Instead, the software routines introduced in this paper automatically selects empirical priors by identifying the hyperparameters $\mu$ and $\sigma^2$ that optimize sensitivity (detecting real effects) and specificity (rejecting absent ones).

Synthetic datasets were generated to mirror the empirical design where half of the effects of the experimental design are switched "on" while the rests remain "off", forming a known ground truth. Noise in the simulated datasets is estimated directly from the empirical data by fitting the design matrix $Z$ to each condition/channel and examining residual variance relative to fitted variance across channels, averaged across subjects, according to:

$$s = \frac{1}{N} \sum_{i=1}^{N} \frac{Var(\boldsymbol{Y_i} - \boldsymbol{XB_i})}{Var(\boldsymbol{XB_i})}$$

Because the optimal prior depends on effect size, our protocol enables one to specify the size of the multivariate ground truth effects (following the implementation described by Lepauvre et al., 2025: https://alexlepauvre.github.io/multisim-neuro/tutorial/07-mathematical_details.html). Effect size can be estimated from previous studies. In our case, we derived effect size from the decoding accuracy observed in (Kaneshiro et al., 2015), under the assumption that their sample was sufficiently powered to approximate the theoretical maximum for a Bayes optimal classifier. The model is then refitted under multiple μ and σ² values (sampled from -64 to -2 and 2 to 64, respectively) to compute each component's free energy. The hyperparameters that provide the best trade-off between positive free energy for present effects and negative free energy for absent effects (rank-sum) are adopted as the final priors for the real data.

Using this pipeline, we obtained an optimal prior expectation of $\mu = -8$ and variance of $\sigma^2 = 4$, based on our empirical data.

## Acknowledgments

Alex Lepauvre is funded by the Templeton World Charity Foundation [TWCF0389–1872 DOI.ORG/10.54224/20389; TWCF0486 - DOI.ORG/10.54224/20486], and the Max Planck Society through Lucia Melloni. Peter Zeidman is funded by a UK Medical Research Council (MRC) Career Development Award [MR/X020274/1]. This research was funded in part by the Discovery Research Platform for Naturalistic Neuroimaging funded by Wellcome [226793/Z/22/Z].

## Authors contributions

**Alex Lepauvre** Conceptualization, Data Curation, Formal Analysis, Methodology, Software, Validation, Visualization, Writing-review & editing. **Lucia Melloni** Conceptualization, Funding Acquisition, Resources, Writing-review & editing. **Karl Friston** Conceptualization, Funding Acquisition, Methodology, Resources, Writing-review & editing. **Peter Zeidman** Conceptualization, Data Curation, Formal Analysis, Funding Acquisition, Methodology, Project Administration, Resources, Software, Supervision, Validation, Visualization, Writing-original draft, Writing-review & editing

## Code availability

All code is available in Github (https://github.com/pzeidman/vRSA_MEEG)

## Declaration of Competing Interests

The authors declare that they have no competing interests.

## Ethics

This study used publicly available, de-identified human data. Therefore, no ethical approval was required.